\documentclass[aps,prl,reprint,groupedaddress,showpacs]{revtex4-1}
\usepackage{graphicx}
\usepackage{dcolumn}
\usepackage{bm}
\usepackage[latin1]{inputenc}
\newcommand{\vE}{\vec{\textbf{\emph{E}}}}

\begin{document}


\title{Clausius-Mossotti Lorentz-Lorenz relations and retardation effects for two-dimensional crystals}

\author{Luca Dell'Anna}
\author{Michele Merano}
\affiliation{Dipartimento di Fisica e Astronomia G. Galilei, Universit$\grave{a}$ degli studi di Padova and CNISM, via Marzolo 8, 35131 Padova, Italy}


\date{\today}

\begin{abstract}
The macroscopic surface electric susceptibility determines the linear optical 
properties of an insulating single-layer two-dimensional atomic crystal, and can be expressed in terms of the microscopic polarizability of the atoms. We compute the local electric field acting on a single atom, both for the static and the dynamic case, as the superposition of the external applied electric field and the fields generated by the induced dipoles in the crystal. We find that, in the dynamic case, retardation effects dephase the local electric field with respect to the incident one. This explains why the Fresnel coefficients of a single-layer two-dimensional atomic crystal are intrinsically complex quantities, even when a null macroscopic surface conductivity is assumed.

\end{abstract}


\maketitle
\section{Introduction}
In 2005 Novoselov and coworkers \cite{Novoselov2005} reported about free-standing atomic crystals, strictly two dimensional (2D) materials, which can be viewed as single atomic planes pulled out of bulk crystals. These atomically thin sheets are stable under ambient conditions, exhibit high crystal quality, and are continuous on a macroscopic scale. Samples of several 2D materials have been obtained and identified, like graphene, hexagonal boron nitride (hBN), $\rm MoS_{2}$ and others. 
They can be conductors (graphene) \cite{Novoselov2004}, semiconductors (transition-metal dichalcogenides) \cite{Heinz2010}, or insulators (hBN) \cite{Blake2011}.

On the macroscopic scale these single-layer 2D atomic crystals can be considered as continuous flat materials, as confirmed by the observations of their optical properties \cite{Nair2008, Blake2007, Kravets2010, Kis11, Heinz2014}. 
It is indeed possible to describe their linear optical response in terms of their surface susceptibility $\chi$ and, for conductors, their surface conductivity $\sigma$ \cite{Merano16, Merano15}. 
These macroscopic quantities can be conceptually introduced without resorting to a microscopic atomic description. It was shown that, as for bulk materials, ellipsometry \cite{Kravets2010} is able to retrieve both $\chi$ and $\sigma$ \cite{Merano16} and that 
the nonlinear optical properties of such 2D atomic crystals \cite{Zhao13, Heinz13, Paula13, Kim14, Cui13, Mikhailov10} can also be described in terms of the macroscopic surface susceptibilities \cite{Merano216}. 

Anyway single-layer atomic crystals hold a surprise. Their Fresnel coefficients are intrinsically complex quantities, even when a null macroscopic surface conductivity is assumed. This is an intriguing dimensionality effect that deserves an explanation. The macroscopic approach presented in Ref.~\cite{Merano16} does not clarify this issue, however it shows that the surface polarization density $\vec{\textbf{\emph{P}}}$ is not in phase with the incident electric field. Since $\vec{\textbf{\emph{P}}}$ is proportional to the local electric field, the latter can not be in phase with the incident electric field \cite{Wolf}.  
This point suggests that the role of the local electric field in a 2D atomic crystal is of great importance.

In this paper we address this issue, by first computing the local electric field in the static case, following the approach developed in Ref.~\cite{Aspnes82}. We then connect the microscopic polarizability to the macroscopic surface susceptibility, obtaining the Clausius-Mossotti expression for single layer 2D atomic crystals. Afterwords we extend our approach to the dynamic case. We then derive the Lorentz-Lorenz expression that relates the dynamic polarizability to the surface susceptibility, which allows us to compute the transmission coefficient and the dephasing of the transmitted electric field with respect to the incident one, finding a perfect agreement with what expected by the macroscopic approach \cite{Merano16}.

\begin{figure}
\includegraphics[width=2.8cm]{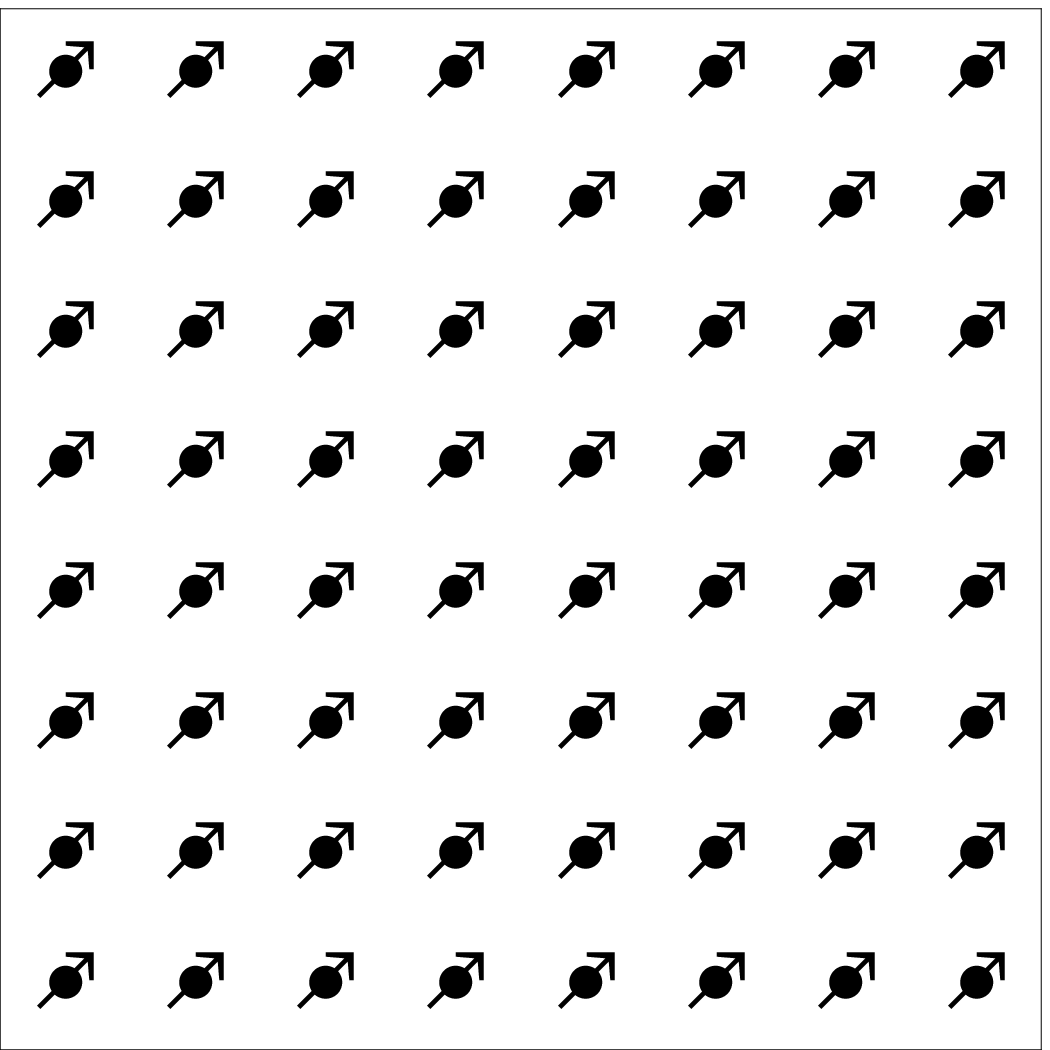}\hspace{0.02cm}
\includegraphics[width=2.8cm]{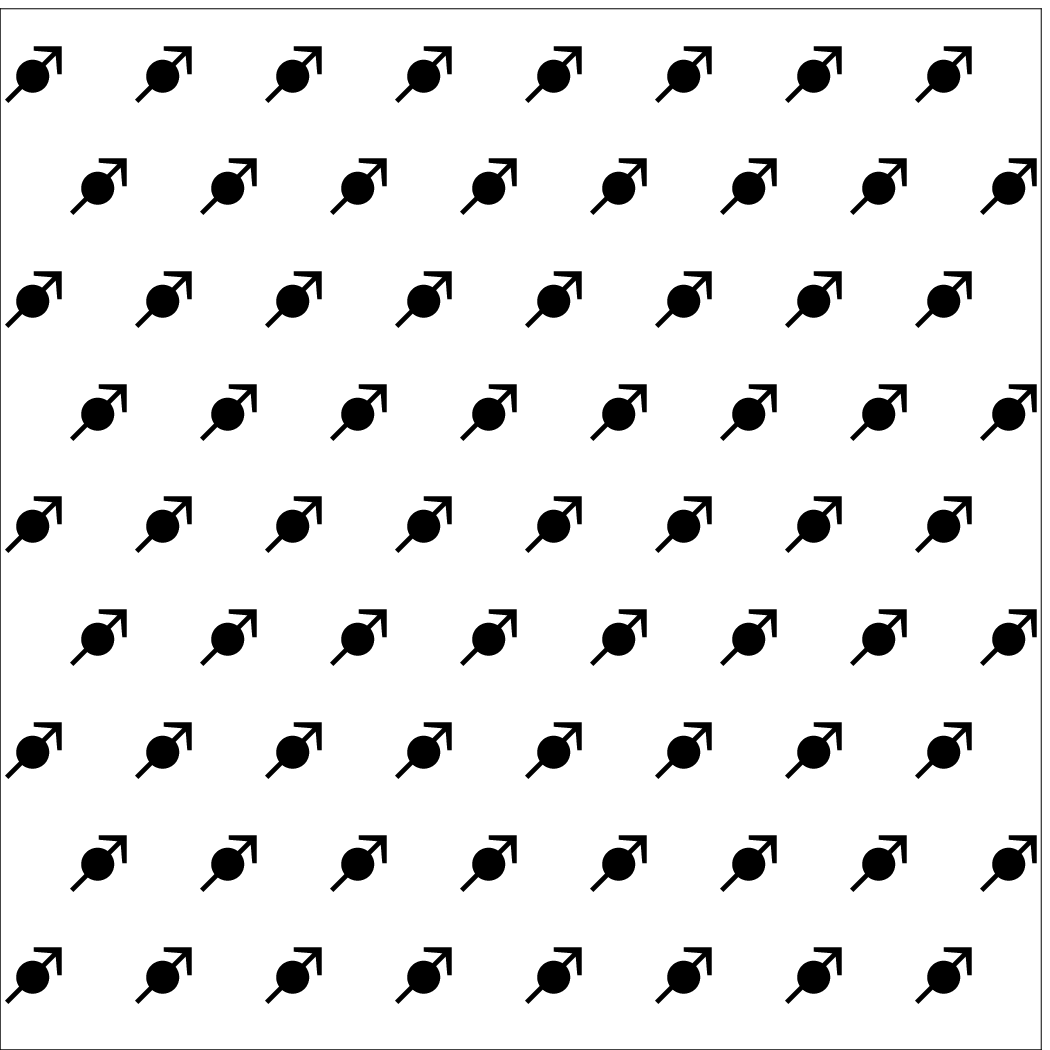}\hspace{0.02cm}
\includegraphics[width=2.8cm]{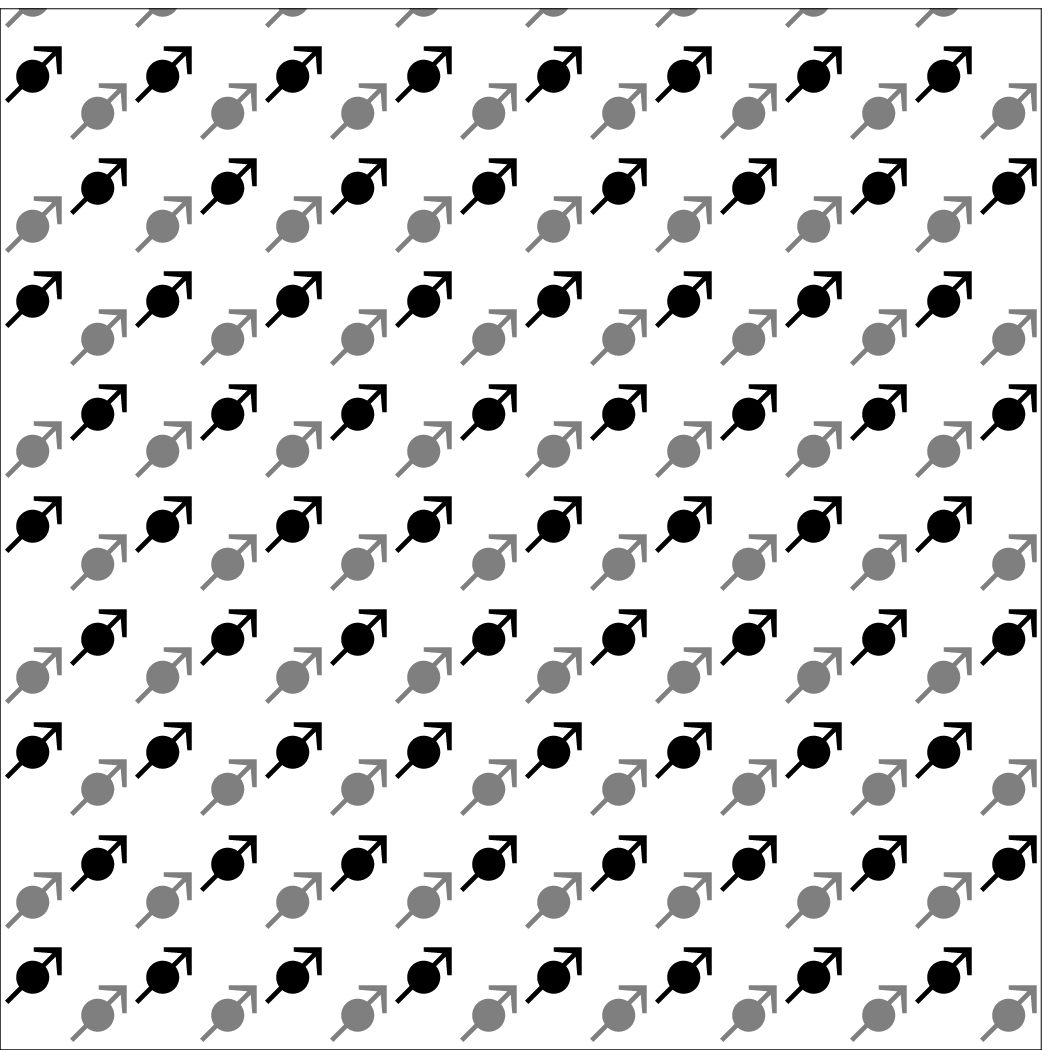}
\caption{Three different dipole distributions, where dipoles are placed on a two-dimensional Bravais lattice, such as a square (left) or a triangular (center) lattice, or on a bipartite lattice, like the honeycomb lattice (right). Dipoles are oriented along the direction of the applied electic field, parallel to the plane.}
\label{fig1}
\end{figure}

\section{Static field}
Let us consider a flat single-layer 2D crystal, composed of atoms with polarizability $\alpha$ \cite{Purcell}. In particular, here and in what follows, we will consider three different lattices: square, triangular and honeycomb lattices (see Fig.~\ref{fig1}.) If we apply an electric field in the plane of the 
crystal a macroscopic dipole moment arises and it is possible to define a polarization density $\vec{\textbf{\emph{P}}}$. If the electric field is 
applied orthogonally to the 2D crystal no macroscopic polarization can be created. Indeed to have a macroscopic polarization the microscopic dipoles need to be aligned, to generate a macroscopic separation of charges.

Let us suppose, therefore, that a static uniform electric field $\vec{\textbf{\emph{E}}}_{i}$ is applied parallel to the plane of a single-layer 2D atomic crystal. The electric field distorts the charge distribution in the crystal, generating electric dipoles, oriented as $\vec{\textbf{\emph{E}}}_{i}$, at the lattice sites. A surface polarization density proportional to the total macroscopic field $\vec{\textbf{\emph{E}}}$ in the crystal may arise:
\begin{eqnarray}
\label{Polarization1}
\vec{\textbf{\emph{P}}}=\epsilon_{0}\chi\vec{\textbf{\emph{E}}}
\end{eqnarray}
 where $\epsilon_{0}$ is the vacuum permittivity and $\chi$ is the electric surface susceptibility and
\begin{eqnarray}
\vec{\textbf{\emph{E}}}=\vec{\textbf{\emph{E}}}_{i}+\vec{\textbf{\emph{E}}}_{p}
\end{eqnarray}
where $\vec{\textbf{\emph{E}}}_{p}$ is the electric field generated by the polarization matter itself \cite{Purcell}.
The polarization density can also be calculated from its definition:
\begin{eqnarray}
\label{Polarization2}
\vec{\textbf{\emph{P}}}=N\vec{\textbf{\emph{p}}}
\end{eqnarray}
where $N$ is the dipole surface density and 
\begin{eqnarray}
\label{Local1}
\vec{\textbf{\emph{p}}}=\alpha \epsilon_{0} \vec{\textbf{\emph{E}}}_{loc}
\end{eqnarray} 
is the induced dipole moment at each reticular point \cite{Aspnes82} and $\vec{\textbf{\emph{E}}}_{loc}$ is the field acting on a single dipole. It is well known that $\vec{\textbf{\emph{E}}}$, $\vec{\textbf{\emph{E}}}_{i}$, $\vec{\textbf{\emph{E}}}_{p}$ and $\vec{\textbf{\emph{E}}}_{loc}$ are in general different quantities. The first three electric fields are macroscopic quantities, the last field is microscopic.
In order to put in relation $\alpha$ with $\chi$, we have to compute the local electric field $\vec{\textbf{\emph{E}}}_{loc}$ and $\vec{\textbf{\emph{E}}}_{p}$. 
\subsection{Local fields}
Let us consider $\vec{\textbf{\emph{E}}}_{loc}$, which can be written as  
\begin{eqnarray}
\label{Local2}
\vec{\textbf{\emph{E}}}_{loc}=\vec{\textbf{\emph{E}}}_{i}
+{\sum_{(m,n)}}'\vec{\textbf{\emph{E}}}_{n,m}
\end{eqnarray}
where the second term on the right-hand side of Eq. (\ref{Local2}) 
is the electric field felt by a dipole on one site, generated by all the other dipoles except the one sitting on that site (the dash in the summation sign indicates that that site is excluded from the sum), and \cite{Purcell}
\begin{eqnarray}
\label{Dipole}
\vec{\textbf{\emph{E}}}_{m,n}=\frac{1}{4\pi \epsilon_{0}}\frac{3(\vec{\textbf{\emph{p}}}\cdot\hat{\textbf{\emph{r}}}_{m,n})\hat{\textbf{\emph{r}}}_{m,n}-\vec{\textbf{\emph{p}}}}{r^3_{m,n}}
\end{eqnarray}
where $\vec{\textbf{\emph{r}}}_{n,m}$ is the vector connecting the origin to the site labelled by the integers ($m, n$). 
$\vec{\textbf{\emph{E}}}_{m,n}$ has a component parallel and a component orthogonal to $\vec{\textbf{\emph{p}}}$. For symmetry reasons only the parallel component gives a contribution to $\vec{\textbf{\emph{E}}}_{loc}$ in Eq.~(\ref{Local2}).

\subsubsection{Square lattice} 

For a square lattice we get
\begin{eqnarray}
\hspace{-0.3cm}
{\sum_{(m,n)}}'\vec{\textbf{\emph{E}}}_{m,n}=\frac{\alpha}{8\pi a^3}
\vec{\textbf{\emph{E}}}_{loc}
\hspace{-0.2cm}{\sum_{(m,n)\neq(0,0)}}
\frac{1}{(n^2+m^2)^{3/2}} 
\end{eqnarray}
where $a$ is the lattice spacing. 
We have therefore that the local field is given by
\begin{eqnarray}
\label{Local3}
\vec{\textbf{\emph{E}}}_{loc}=\vec{\textbf{\emph{E}}}_{i}+\frac{\alpha \,C_0}
{4\pi a^3}\vec{\textbf{\emph{E}}}_{loc}
\end{eqnarray}
where $C_0$, obtained by the sum over all couples of integers $(m,n)$ except $(0,0)$ is 
\begin{eqnarray}
\label{cos}
C_0=2\,\zeta\Big(\frac{3}{2}\Big)\,L\Big(\frac{3}{2},{\cal X}_4\Big)=4.51681...
\end{eqnarray}
with $\zeta(s)=\sum^\infty_{n=1}\frac{1}{n^s}$ the Rienman zeta function
and $L(s,{\cal X}_4)=\sum_{n=1}^\infty \frac{{\cal X}_4(n)}{n^s}$
a Dirichlet L-function, 
with ${\cal X}$ a Dirichlet character to the modulus $4$: ${\cal X}_4(0)=0$, ${\cal X}_4(1)=1$, ${\cal X}_4(2)=0$, ${\cal X}_2(3)=-1$, and  ${\cal X}_4(n+4)={\cal X}_4(n)$, therefore $L(s,{\cal X}_4)\equiv\beta(s)=\sum^\infty_{n=0}\frac{1}{(2n+1)^s}$ is the so-called Dirichlet beta-function.

\subsubsection{Triangular lattice}

For a triangular lattice
\begin{eqnarray}
\hspace{-0.3cm}
{\sum_{(m,n)}}'\vec{\textbf{\emph{E}}}_{m,n}=\frac{\alpha}{8\pi a^3}
\vec{\textbf{\emph{E}}}_{loc}
\hspace{-0.4cm}{\sum_{(m,n)\neq(0,0)}}\hspace{-0.05cm}
\frac{1}{(n^2+nm+m^2)^{3/2}}
\end{eqnarray}
which, inserted into Eq.~(\ref{Local2}) yields Eq.~(\ref{Local3}) with
\begin{equation}
\label{cot}
C_0=3\,\zeta\Big(\frac{3}{2}\Big)\,L\Big(\frac{3}{2}, {\cal X}_3\Big)=5.51709...
\end{equation}
where $L(s,{\cal X}_3)=\sum_{n=1}^\infty \frac{{\cal X}_3(n)}{n^s}$ is another Dirichlet L-series with modulus $3$, where ${\cal X}_3(0)=0$, ${\cal X}_3(1)=1$, ${\cal X}_3(2)=-1$, and ${\cal X}_3(n+3)={\cal X}_3(n)$. 

\subsubsection{Honeycomb lattice}

This is a bipartite lattice made of two copies of triangular lattices, that we label $1$ and $2$. For the sake of generality we can suppose to have two different polarizabilities for the dipoles sitting on the two triangular lattices. We can still define a macroscopic polarization density as
\begin{equation}
\label{Phoney}
\vec{\textbf{\emph{P}}}=\frac{N}{2}(\vec{\textbf{\emph{p}}}_1+\vec{\textbf{\emph{p}}}_2)
\equiv \frac{N\epsilon_0}{2}\left(\alpha_1 \vec{\textbf{\emph{E}}}^{(1)}_{loc}+\alpha_2\vec{\textbf{\emph{E}}}^{(2)}_{loc}\right)
\end{equation}
where $N$ is the density of dipoles on the honeycomb lattice ($N=4/(\sqrt{3}a^2)$, in terms of the triangular lattice spacing) and
\begin{equation}
\vec{\textbf{\emph{E}}}^{(\ell)}_{loc}=\vec{\textbf{\emph{E}}}_i+
{{\sum_{(m,n)}}'}^{(\ell)}\vec{\textbf{\emph{E}}}_{n,m}
\end{equation}
with $\ell=1,2$, the local fields on the two different sublattices, and 
\begin{eqnarray}
\nonumber{{\sum_{(m,n)}}'}^{(1)}\vec{\textbf{\emph{E}}}_{n,m}&=&
\frac{1}{4\pi \epsilon_{0}}\left({\sum_{(m,n)\neq(0,0)}}
\frac{3(\vec{\textbf{\emph{p}}}_1
\cdot\hat{\textbf{\emph{r}}}_{m,n})\hat{\textbf{\emph{r}}}_{m,n}-
\vec{\textbf{\emph{p}}}_1}{r^3_{m,n}}\right.\\
&&\left.+{\sum_{(m,n)}}\frac{3(\vec{\textbf{\emph{p}}}_2
\cdot\hat{\textbf{\emph{r}}}'_{m,n})\hat{\textbf{\emph{r}}}'_{m,n}-
\vec{\textbf{\emph{p}}}_2}{{r'}^3_{m,n}}\right)\\
\nonumber{{\sum_{(m,n)}}'}^{(2)}\vec{\textbf{\emph{E}}}_{n,m}&=&
\frac{1}{4\pi \epsilon_{0}}\left({\sum_{(m,n)\neq(0,0)}}
\frac{3(\vec{\textbf{\emph{p}}}_2
\cdot\hat{\textbf{\emph{r}}}_{m,n})\hat{\textbf{\emph{r}}}_{m,n}-
\vec{\textbf{\emph{p}}}_2}{r^3_{m,n}}\right.\\
&&\left.+{\sum_{(m,n)}}\frac{3(\vec{\textbf{\emph{p}}}_1
\cdot\hat{\textbf{\emph{r}}}'_{m,n})\hat{\textbf{\emph{r}}}'_{m,n}-
\vec{\textbf{\emph{p}}}_1}{{r'}^3_{m,n}}\right)
\end{eqnarray}
where $\hat{\textbf{\emph{r}}}_{m,n}$ are the Bravais vectors and $\hat{\textbf{\emph{r}}}'_{m,n}$ the vectors defining the second sublattice. 
After defining 
\begin{eqnarray}
\label{coe1}
C_0^{(1)}&=&\hspace{-0.2cm}\sum_{(m,n)\neq(0,0)}\frac{1}{2(n^2+nm+m^2)^{\frac{3}{2}}}=5.51709...\\
\nonumber C_0^{(2)}&=&\sum_{(m,n)}\frac{1}{2(n^2+nm+m^2+n+m+\frac{1}{3})^{\frac{3}{2}}}\\
&=&11.5753...
\label{coe2}
\end{eqnarray}
we obtain for the local fields
\begin{eqnarray}
\label{Localexagon}
\vec{\textbf{\emph{E}}}^{(1)}_{loc}=\vec{\textbf{\emph{E}}}_{i}+
\frac{\alpha_1 \,C^{(1)}_0}{4\pi a^3}
\vec{\textbf{\emph{E}}}^{(1)}_{loc}+
\frac{\alpha_2 \,C^{(2)}_0}{4\pi a^3}
\vec{\textbf{\emph{E}}}^{(2)}_{loc}\\
\vec{\textbf{\emph{E}}}^{(2)}_{loc}=\vec{\textbf{\emph{E}}}_{i}+
\frac{\alpha_2 \,C^{(1)}_0}{4\pi a^3}\vec{\textbf{\emph{E}}}^{(2)}_{loc}+
\frac{\alpha_1 \,C^{(2)}_0}{4\pi a^3}\vec{\textbf{\emph{E}}}^{(1)}_{loc}
\label{Localexagon2}
\end{eqnarray}
where $a$ is the triangular Bravais lattice spacing.

\subsection{Clausius-Mossotti formula}

Let us consider $\vec{\textbf{\emph{E}}}_{p}$. The static and 
uniform $\vec{\textbf{\emph{E}}}_{i}$ generates an in-plane uniform planar polarization density 
$\vec{\textbf{\emph{P}}}(x, y, z)=\vec{\textbf{\emph{P}}}_{0}
\delta (z)$, consequently $\vec{\textbf{\emph{E}}}_{p}$ is given by \cite{Stratton}:
\begin{eqnarray}
\vec{\textbf{\emph{E}}}_{p}&=&\nabla (\nabla \cdot \vec{\bf{{\Pi}}})\\
\vec{\bf{{\Pi}}}(x', y', z')&=&\frac{1}{4\pi \epsilon_{0}}\int\frac{\vec{\textbf{\emph{P}}}(x, y, z)}{r}dv 
\end{eqnarray}
where $\vec{\bf{\Pi}}$ is the polarization potential. Performing the calculation we find that  $\vec{\textbf{\emph{E}}}_{p}$ is a null vector. 


At this point we are able to compute the Clausius-Mossotti formula for a 
single-layer 2D atomic crystal. 

For the square and the triangular lattice, 
from Eqs.~(\ref{Polarization1}), (\ref{Polarization2}), (\ref{Local1}) and 
(\ref{Local3}) we have:
\begin{eqnarray}
\chi=\frac{N\alpha}{1-\frac{C_0\alpha}{4\pi a^3}}
\label{chi}
\end{eqnarray}
Expressing $a$ in terms of $N$ ($N=1/a^2$ for square and $N=2/(\sqrt{3}a^2)$ 
for triangular lattice) we obtain:
\begin{eqnarray}
\label{chitilde}
\chi= \frac{N\alpha}{1-\widetilde C_0\,N^{3/2}\alpha}
\end{eqnarray}
with $\widetilde C_0\simeq 5.51709(\sqrt{3}/2)^{3/2}/4\pi \simeq 0.3538$ for 
triangular lattice and $\widetilde C_0\simeq 4.5168/4\pi\simeq 0.3594$ 
for square lattice. 

For the honeycomb lattice, from Eqs.~(\ref{Polarization1}), (\ref{Phoney}), (\ref{Localexagon}) and (\ref{Localexagon2}), we obtain
\begin{equation}
\label{chihoney}
\chi=\frac{N}{2}\frac{\alpha_1+\alpha_2-\frac{2\alpha_1\alpha_2\left(C_0^{(1)}-C_0^{(2)}\right)}{4\pi a^3}}{1-\frac{C_0^{(1)}(\alpha_1+\alpha_2)}{4\pi a^3}+\frac{\alpha_1\alpha_2\left(C_0^{(1)2}-C_0^{(2)2}\right)}{(4\pi a^3)^2}}
\end{equation}
Expressing $a$ in terms of the density for the honeycomb lattice $N={4}/{\sqrt{3}a^2}$, Eq.~(\ref{chihoney}) can be written as follows
\begin{equation}
\chi=\frac{N\alpha_{a}}{\frac{1+\Delta_0 N^{3/2}\alpha_a}{1+\Delta_0 N^{3/2}\alpha_h}-\widetilde C_0 N^{3/2}\alpha_a}
\label{chihoney2}
\end{equation}
where $\widetilde C_0=(C_0^{(1)}+C_0^{(2)})(\sqrt{3}/4)^{3/2}/4\pi\simeq 0.38756$ and $\Delta_0=(C_0^{(2)}-C_0^{(1)})(\sqrt{3}/4)^{3/2}/4\pi\simeq 0.13737$, 
while $\alpha_a=(\alpha_1+\alpha_2)/2$ is the arithmetic mean and $\alpha_h=2/(\alpha_1^{-1}+\alpha_2^{-1})$ the harmonic mean.\\
In the special case $\alpha_{1}=\alpha_{2}$, Eq.~(\ref{chihoney2}) reduces to 
Eq.~(\ref{chitilde}).  
Notice that Eq.~(\ref{chihoney2}) is valid for any bipartite lattice. 
 
These results have to be contrasted with the standard Clausius-Mossotti relation in $3$ dimensions (where $\vec{\textbf{\emph{E}}}_{p}=-\vec{\textbf{\emph{P}}}/3\epsilon_0$, and $\vec{\textbf{\emph{E}}}_{i}=\vec{\textbf{\emph{E}}}_{loc}$) which, by the same notation, reads 
\begin{equation}
\chi_{3D}= \frac{N\alpha}{1-\frac{1}{3}N\alpha}.
\end{equation}
where here $N$ is the volume dipole density in three dimensions. It is remarkable that, in contrast to the three dimensional case, the Clausius and Mossotti formula in two dimensions depends on the underlying lattice. 

\section{Dynamic field: the role of retardation}

We consider now a plane wave incident on a single-layer 2D atomic crystal. For simplicity we assume normal incidence. In the plane of the crystal this field has a time dependance given by:
\begin{eqnarray}
\vec{\textbf{\emph{E}}}_{i}(t)=\vec{\textbf{\emph{E}}}_{i}\,e^{i\omega t}
\end{eqnarray}
Again a surface polarization arises
\begin{eqnarray}
\label{Polarization1t}
\vec{\textbf{\emph{P}}}(t)=\epsilon_{0}\chi\vec{\textbf{\emph{E}}}(t)
\end{eqnarray}
where the total macroscopic electric field $\vec{\textbf{\emph{E}}}(t)$ in the crystal is given by \cite{Merano16}
\begin{eqnarray}
\label{Merano}
\vec{\textbf{\emph{E}}}(t)=\vec{\textbf{\emph{E}}}_{\rm t}(t)=\textrm{t}\vec{\textbf{\emph{E}}}_{i}(t).
\end{eqnarray}
Here $\vec{\textbf{\emph{E}}}_{\rm{t}}$ is the transmitted electric field and $\rm t$ is the complex transmission coefficient \cite{Merano16}. Equations (\ref{Polarization2}) and (\ref{Local1}) relating $\vec{\textbf{\emph{P}}}(t)$ 
and $\vec{\textbf{\emph{E}}}_{loc}(t)$ are still valid. We make the assumption that
\begin{eqnarray}
\vec{\textbf{\emph{E}}}_{loc}(t)=\vec{\textbf{\emph{E}}}_{loc}\,e^{i(\omega t+\varphi)}
\end{eqnarray}
where $\vec{\textbf{\emph{E}}}_{loc}(t)$ and hence $\vec{\textbf{\emph{P}}}(t)$ have the same frequency of $\vec{\textbf{\emph{E}}}_{i}(t)$ but eventually a different phase.

\subsection{Local fields}

Let us now compute $\vec{\textbf{\emph{E}}}_{loc}$, the electric field felt by a single dipole, which can be written as 
the sum of the incident electric field $\vec{\textbf{\emph{E}}}_{i}(t)$ and 
the fields of all the other dipoles, as in Eq.~(\ref{Local2}). In the dynamic case the expression for $\vec{\textbf{\emph{E}}}_{m,n}(t)$ has a different dependance on $\vec{\textbf{\emph{p}}}(t)$ with respect to the static case \cite{Wolf, Feynman, Jackson}: 
\begin{eqnarray}
\label{Dipole2}
\vec{\textbf{\emph{E}}}_{m,n}(t)&=&\frac{1}{4\pi \epsilon_{0}r^3_{m,n}} \biggl(3(\tilde{\vec{\textbf{\emph{p}}}}\cdot\hat{\textbf{\emph{r}}}_{m,n})\hat{\textbf{\emph{r}}}_{m,n}-\tilde{\vec{\textbf{\emph{p}}}} \nonumber \\
&-&\frac{1}{c^2}(\hat{\textbf{\emph{r}}}_{m,n} \times \ddot{\vec{\textbf{\emph{p}}}}) \times \hat{\textbf{\emph{r}}}_{m,n} \biggr)
\end{eqnarray}
where 
\begin{eqnarray}
\tilde{\vec{\textbf{\emph{p}}}}=\vec{\textbf{\emph{p}}}(t-\frac{r}{c})+\frac{r}{c}\dot{\vec{\textbf{\emph{p}}}}(t-\frac{r}{c})
\end{eqnarray}

As in the static case, only the component of $\vec{\textbf{\emph{E}}}_{m,n}(t)$ parallel to $\vec{\textbf{\emph{p}}}$ contribute to $\vec{\textbf{\emph{E}}}_{loc}(t)$ and, contrary to the 3-dimensional case its expression depends on the lattice. 

\subsubsection{Square and triangular lattices}
Both for the square and the triangular lattice, the sum of Eq.(\ref{Dipole2}) 
over the sites can be expressed only in terms of the distances $r_{m,n}$
\begin{eqnarray}
\label{STt}
\nonumber
\sum'_{(m,n)}&&\vE_{m,n}(t)= 
\frac{\alpha}{4\pi}\vE_{loc}(t)\\
&&\times\sum'_{(m,n)}\left\{e^{-ikr_{m,n}}
\left(
\frac{1+ik\,r_{m,n}+k^2r_{m,n}^2}{2\,r_{m,n}^{3}}\right)\right\}
\label{sumq}
\end{eqnarray}
where $k=\omega/c=2\pi/\lambda$, being $\lambda$ the wavelength of 
the incident wave, 
\begin{eqnarray}
r_{m,n}=a\sqrt{n^2+m^2}
\label{rsl}
\end{eqnarray}
for the square lattice, and 
\begin{eqnarray}                                                               
r_{m,n}=a\sqrt{n^2+nm+m^2}                                                     
\label{rtl}                                                                    
\end{eqnarray}                                                                 
for the triangular lattice.

\begin{figure}
\includegraphics[width=7.5cm]{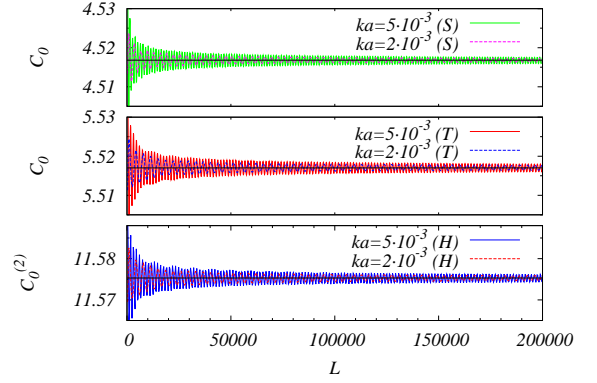}
\caption{$C_0$, as a function of the lattice size $L$ ($-L\le n,m\le L$),
nuerically obtained from the real part of the r.h.s. of
Eq.~(\ref{sumq}), for two different values of $ka$, for square lattice
(S, top panel), triangular lattice (T, middle panel). $C_0^{(2)}$ (H, bottom panel), for honeycomb lattice, is also shown, numerically obtained from the second term in the r.h.s. of Eq.~(\ref{sumh}) 
(while $C_0^{(1)}$ is equal to $C_0$ of the triangular lattice.)
A slow convergence to the static values Eqs.~(\ref{cos}), (\ref{cot}), (\ref{coe2}) (black straight lines) is observed.
The two wavelengths considered, for a typical value of $a\simeq 2.5{\textrm{\AA}}$, are in the range of the visible spectrum.}
\label{fig2}
\end{figure}

\begin{figure}
\noindent\includegraphics[width=7.5cm]{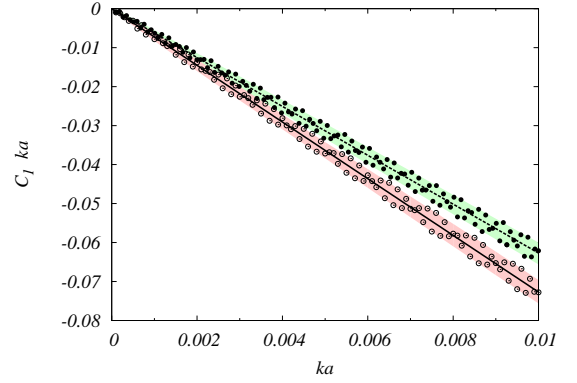}
\caption{$C_1ka$ as a function of $ka$, for triangular lattice (circles) and 
square lattice (dots), with the best fits for the triangular (solid line) and 
square (dashed line) lattices. The shadow regions show the uncertainty of the 
slop due to the finite size over which the sums are performed: $(2L)^2$ is the 
number of sites, with $L=32000$ for the square lattice and $L=40000$ for the 
triangular lattice. This figure shows that, within the numerical error, $C_1$ does not depend on $a$ and $k$.}
\label{fig3}
\end{figure}

\begin{figure}
\includegraphics[width=7.5cm]{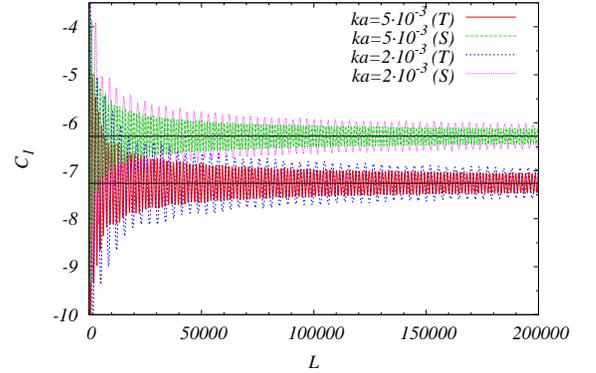}
\caption{$C_1$ as a function of the lattice size $L$ ($-L\le n,m\le L$),
nuerically obtained from the imaginary part of the r.h.s. of
Eq.~(\ref{sumq}), for two different values of $ka$, for square lattice
(S, top curves) and triangular lattice (T, bottom curves). 
$C_1$ for the honeycomb lattice is equal to the one of the triangular lattice.
The numerical values of $C_1$ slowly convergence to the values in Eqs.~(\ref{c1q}) and (\ref{c1t}) (black straight lines).
}
\label{fig4}
\end{figure}

Contrary to the static case we are not able to analytically solve Eq.~(\ref{STt}). By numerical summation we obtain
\begin{eqnarray}
{\sum_{(m,n)}}'\vec{\textbf{\emph{E}}}_{m,n}(t)=\frac{\alpha}{4\pi a^3}
\big(C_0+i\, C_1 k a\big)
\vec{\textbf{\emph{E}}}_{loc}(t)
\end{eqnarray}
where the real part of the summation is proportional to $C_0$ and it converges to the same values (Eqs. (\ref{cos}) and (\ref{cot})) obtained for the static fields, while the imaginary part is proportinal to $C_1$ where
\begin{equation}
\label{c1q}
C_1\simeq -6.28
\end{equation}
for the square lattice and
\begin{equation}
\label{c1t}
C_1\simeq -7.26
\end{equation}
for the triangular one.
Actually those values of $C_0$ and $C_1$ are reached when the sum in 
Eq.~(\ref{sumq}) is extended to a very large number of sites, as shown in 
Figs.~\ref{fig2}, \ref{fig3}, \ref{fig4}. We stress that both $C_0$ and $C_1$ are constant numbers that do not depend on $k$ and $a$.

The local electric field is, then, connected to the incident one through the following equation
\begin{eqnarray}
\label{Local3d}
\vec{\textbf{\emph{E}}}_{loc}
\left(1-\frac{\alpha C_0}{4\pi a^3}-i \frac{\alpha C_1 k}{4\pi a^2}\right)
e^{i\varphi}=\vec{\textbf{\emph{E}}}_{i}
\end{eqnarray}

\subsubsection{Honeycomb lattice}
As already said, for a bipartite lattice we have to distinguish two
local fields felt by the dipoles sitting on the two different sublattices.
For the honeycomb lattice we have, therefore,
\begin{eqnarray}
\label{sumh}
&&{{\sum_{(m,n)}}'}^{(1)}\vE_{m,n}(t)=
\frac{1}{4\pi}\\
\nonumber&&\hspace{-0.cm}\times\left\{\alpha_1\vE_{loc}^{(1)}(t)
\hspace{-0.3cm}\sum_{(m,n)\neq(0,0)}\left[e^{-ikr_{m,n}}
\left(\frac{1+ik\,r_{m,n}+k^2r_{m,n}^2}{2\,r_{m,n}^{3}}\right)\right]\right.\\
\nonumber&&\hspace{-0.cm}+\left.
\alpha_2\vE_{loc}^{(2)}(t)\sum_{(m,n)}\left[e^{-ikr'_{m,n}}
\left(\frac{1+ik\,r'_{m,n}+k^2{r'}_{m,n}^{2}}{2\,{r'}_{m,n}^{3}}\right)\right]
\right\}
\end{eqnarray}
and an analogous expression for $\sum_{(m,n)}^{'(2)}\vE_{m,n}(t)$ where $\alpha_1\vE_{loc}^{(1)}(t)$ and $\alpha_2\vE_{loc}^{(2)}(t)$ are exchanged, and where
\begin{eqnarray}
&&r_{m,n}=a\sqrt{n^2+nm+m^2}\\
&&r'_{m,n}=a\sqrt{n^2+nm+m^2+n+m+{1}/{3}}
\end{eqnarray}
As a result, the local fields are defined by 
\begin{eqnarray}
\label{Localhoney1}
\nonumber\vec{\textbf{\emph{E}}}_{i}(t)&=&\vec{\textbf{\emph{E}}}^{(1)}_{loc}(t)
\left(1-\frac{\alpha_1 C_0^{(1)}}{4\pi a^3}-i \frac{\alpha_1 C_1 k}{4\pi a^2}
\right)\\
&&-\vec{\textbf{\emph{E}}}^{(2)}_{loc}(t)\left(\frac{\alpha_2 C_0^{(2)}}
{4\pi a^3}+i \frac{\alpha_2 C_1 k}{4\pi a^2}\right)\\
\nonumber\vec{\textbf{\emph{E}}}_{i}(t)&=&\vec{\textbf{\emph{E}}}^{(2)}_{loc}(t)
\left(1-\frac{\alpha_2 C_0^{(1)}}{4\pi a^3}-i \frac{\alpha_2 C_1 k}{4\pi a^2}
\right)\\
&&-\vec{\textbf{\emph{E}}}^{(1)}_{loc}(t)\left(\frac{\alpha_1 C_0^{(2)}}
{4\pi a^3}+i \frac{\alpha_1 C_1 k}{4\pi a^2}\right)
\label{Localhoney2}
\end{eqnarray}
where $C_0^{(1)}$ and $C_0^{(2)}$ are given by
Eqs.~(\ref{coe1}),~(\ref{coe2}) and $C_1$ by Eq.~(\ref{c1t}). Notice that $C_0^{(1)}$ and $C_1$ are equal to those for the triangular lattice.



\subsection{Lorentz-Lorenz formula and transmission coefficient}

Following the same reasoning of the static case we show now that we can fix 
both the Lorentz-Lorenz formula and the Fresnel coefficients for a single-layer 2D atomic crystal. 

Let us start considering the square and the triangular lattices. Since $\vec{\textbf{\emph{E}}}_{i}$ and $\vec{\textbf{\emph{E}}}_{loc}$ are fixed to be real, from (\ref{Local3d}) we have
\begin{eqnarray}
\tan(\varphi)=\frac{\frac{\alpha C_1 k}{4\pi a^2}}{1-\frac{\alpha C_0}{4\pi a^3}}
\label{tan}
\end{eqnarray}
%
The relation between $\chi$ and $\alpha$ is obtained from Eqs.~(\ref{Polarization2}), (\ref{Local1}), (\ref{Polarization1t}), (\ref{Merano}), and (\ref{Local3d}):
\begin{eqnarray}
\label{eqfinal}
N\alpha=\chi \left|\rm t\right|\sqrt{\left(1-\frac{\alpha C_0}{4\pi a^3}\right)^2+\left(\frac{\alpha C_1 k}{4\pi a^2}\right)^2}
\end{eqnarray}
which can be written as
\begin{eqnarray}
\label{modt}
\frac{N\alpha}{1-\frac{\alpha C_0}{4\pi a^3}}=\chi \left|\rm t\right|
\sqrt{1+\tan(\varphi)^2}
\end{eqnarray}
Since, for $k=0$ we should have that $|\textrm{t}|=1$ and $\tan(\varphi)=0$, 
then 
$\chi$ has to be equal to the static result, Eq. (\ref{chi}), therefore
\begin{equation}
|\textrm{t}|=\frac{1}{\sqrt{1+\tan(\varphi)^2}}
\end{equation}
Using again Eq. (\ref{chi}), we can rewrite Eq. (\ref{tan})
\begin{equation}
\label{tanphi}
\tan(\varphi)=\frac{C_1}{2\pi N a^2}\frac{k\,\chi}{2}=-\frac{k\chi}{2}
\end{equation}
since, both for a square lattice, where $N=1/a^2$ and $C_1$ is compatible with 
$C_1= -2\pi\simeq -6.28$ 
(see Eq.(\ref{c1q}), and for the triangular lattice, 
where $N=2/\sqrt{3}a^2$ and $C_1= -4\pi/\sqrt{3}\simeq -7.26$ (see Eq.(\ref{c1t})), we have 
\begin{equation}
\label{c1general}
{C_1}= -2\pi Na^2\,.
\end{equation}
The value of $|\textrm{t}|$ and $\tan(\varphi)$ for perpendicularly incidet electromagnetic wave is in perfect agreement with the solution of the field equations with the proper boundary conditions \cite{Merano16}. 

In the case of a honeycomb lattice, using Eqs.~(\ref{Phoney}), (\ref{Polarization1t}), (\ref{Merano}), (\ref{Localhoney1}), (\ref{Localhoney2}) 
we get
\begin{widetext}
\begin{eqnarray}
\label{eqfinal2}
\chi \,{\rm t}=\frac{N}{2}\frac{\alpha_1+\alpha_2-\frac{2\alpha_1\alpha_2\left(C_0^{(1)}-C_0^{(2)}\right)}{4\pi a^3}}{1-\frac{C_0^{(1)}(\alpha_1+\alpha_2)}{4\pi a^3}+
\frac{\alpha_1\alpha_2\left(C_0^{(1)2}-C_0^{(2)2}\right)}{(4\pi a^3)^2}
-i C_1k\frac{1}{4\pi a^2}\left[\alpha_1+\alpha_2-\frac{2\alpha_1\alpha_2\left(C_0^{(1)}
-C_0^{(2)}\right)}{4\pi a^3}\right]}
\end{eqnarray}
\end{widetext}
Using Eq.~(\ref{chihoney}) one can easily check that 
\begin{equation}
{\rm t}=\frac{1}{1-i\frac{C_1 k}{2\pi N a^2}\chi}=\frac{1}{1+i\frac{\chi k}
{2}}
\end{equation}
where we used $C_1=-4\pi/\sqrt{3}\simeq -7.26$, Eq.~(\ref{c1t}), and $N=4/(\sqrt{3}a^2)$, the density for the honeycomb lattice. 

We have shown, therefore, that for all the lattices considered,  
the transmission coefficient depends only on the electric susceptibility $\chi$, in agreement with the macroscopic apprach \cite{Merano16}. $\chi$, in its turn, depends on the geometry of the underlying lattices.

\section{Conclusions}

We derived the Clausius-Mossotti Lorentz-Lorenz relations for single-layer 
two-dimensional atomic crystals. In contrast to the three dimensional case, 
these expressions depend on the underlying atomic lattices, due to the local 
electric field acting on the single dipoles.

In three-dimensional crystals, for static field, the local field is simply equal to the applied external electric field \cite{Aspnes82}. In the dynamic case, the Ewald and Oseen theorem \cite{Wolf} explains how, through the interference with the retarded dipole electric fields, the incident wave is replaced by a polarization wave which acts on the atoms of the crystal and propagates with a velocity which is smaller than that of the electromagnetic field in the vacuum. This theorem, when applied to the case of a monochromatic plane-wave entering from vacuum into a dielectric transparent medium, leads to reflection and refraction laws and to real Fresnel coefficients. 
    
In two-dimensional crystals, instead, all the dipoles contribute to the local electric field both in the static and in the dynamic cases. 
In the dynamic case, due to the finite velocity of propagation of the dipolar potential, a dephasing between the local field, and the incident electric field is induced. On the macroscopic scale, this translates to a dephasing between the incident electric field and the polarization density, which is the origin of intrinsically complex Fresnel coefficients even for null surface conductivity. 
This dephasing, due to low dimensionality, is a direct manifestation of retardation effects in the linear optical response of single-layer 2D atomic crystals.

\acknowledgements{L.D. acknowledges financial support from MIUR, through FIRB Project No. RBFR12NLNA\_002, and PRIN Project 2010LLKJBX.}
\bibliography{letter}
\end{document}